\documentclass{PoS}

\usepackage{amsfonts} 
\usepackage{amssymb} 
\usepackage{amsmath} 
\usepackage{subfigure}
\usepackage{graphicx} 
\usepackage{array} 
\usepackage{dcolumn} 
\usepackage{bm} 
\usepackage{latexsym} 
\usepackage{longtable} 
\usepackage{color}
\usepackage{mathrsfs}

\graphicspath{{./figs/}}

\newcommand{\GeV}{\mathop{\rm GeV}\nolimits}

\title{ 
  Beyond the Standard Model corrections to $K^0-\bar{K}^0$ mixing 
}

\ShortTitle{
  BSM corrections to $K^0-\bar{K}^0$ mixing
}

\author{\speaker{Hyung-Jin Kim}, Chulwoo Jung \\
  Physics Department, Brookhaven National Laboratory,
  Upton, NY11973, USA \\
  E-mail: \email{hjkim@bnl.gov}}

\author{Jon A. Bailey, Yong-Chull Jang, Hwancheol Jeong,
  Jangho Kim, Kwangwoo Kim, Seonghee Kim, Jaehoon Leem, Boram Yoon,
  Weonjong Lee, \\
  Lattice Gauge Theory Research Center, CTP, and FPRD, \\
  Department of Physics and Astronomy,
  Seoul National University, Seoul, 151-747, South Korea \\
  E-mail: \email{wlee@snu.ac.kr}}

\author{Taegil Bae \\
  Korea Institute of Science and Technology Information,
  Daejeon, 305-806, South Korea \\
  E-mail: \email{esrevinu@gmail.com}}

\author{Jongjeong Kim \\
  Physics Department,
  University of Arizona,
  Tucson, AZ 85721, USA \\
  E-mail: \email{rvanguard@gmail.com}}

\author{Stephen R. Sharpe\\
  Physics Department, University of Washington, 
  Seattle, WA 98195-1560, USA \\
  E-mail: \email{sharpe@phys.washington.edu}}

\author{SWME Collaboration}

\abstract{
  We calculate the B-parameters for operators arising in theories
  of new physics beyond the standard model (BSM)
  using HYP-smeared improved staggered fermions on the MILC asqtad 
  lattices with $N_f = 2+1$ flavors.
  We use three different lattice spacings ($a \approx 0.045,~ 0.06$ 
  and $0.09 \text{ fm}$) at a fixed ratio of light to strange
  quarks, $m_\ell/m_s = 1/5$, 
  to obtain the continuum results.
  Operator matching is done using perturbative matching at one-loop
  order, and results are run to 2 or 3 GeV using two-loop running
  in the $\overline{\rm MS}$ scheme.
  For the chiral and continuum extrapolations, we use SU(2)
  staggered chiral perturbation theory.
  We present preliminary results with only statistical errors.
}

\FullConference{
The 30th International Symposium on Lattice Field Theory - Lattice 2012 \\
June 24 -- 29, 2012\\
Cairns, Australia
}

\begin{document}

\section{Introduction} 

In the standard model, mixing in the neutral kaon system arises due to the 
weak interaction.
Integrating out the heavy particles, the mixing is described by the
matrix element of a $\Delta S = 2$ four-fermion operator
($Q^\text{Cont}_1$ below), and is parametrized by $B_K$.
$B_K$ is now determined with high precision from lattice QCD
\cite{Aubin:2009jh,Aoki:2010pe,Durr:2011ap,Bae:2011ff}, and plays
an important role in constraining the parameters of the CKM matrix.
In BSM theories, additional operators contribute to kaon mixing. 
If the matrix elements of these operators were known, one could
constrain the parameters of these theories in a way that
is complementary to direct searches.
Here we present a calculation of the new matrix elements using
HYP-smeared staggered valence fermions on the 
MILC asqtad lattices.

We adopt the operator basis used in perturbative
calculations of anomalous dimensions~\cite{Buras:2000if}
\begin{align}
 \label{eq:op_bk}
 {Q}^\text{Cont}_{1} &=
 [\bar{s}^a \gamma_\mu (1-\gamma_5) d^a] 
 [\bar{s}^b \gamma_\mu (1-\gamma_5) d^b],   \\
 \label{eq:op_b2}
 {Q}^\text{Cont}_{2} &=
 [\bar{s}^a (1-\gamma_5) d^a] [\bar{s}^b (1-\gamma_5) d^b],   \\
 \label{eq:op_b3}
 {Q}^\text{Cont}_{3} &=
 [\bar{s}^a \sigma_{\mu\nu}(1-\gamma_5) d^a] 
 [\bar{s}^b \sigma_{\mu\nu} (1-\gamma_5) d^b],   \\
 \label{eq:op_b4}
 {Q}^\text{Cont}_{4} &=
 [\bar{s}^a (1-\gamma_5) d^a] 
  [\bar{s}^b (1+\gamma_5) d^b],  \\
 \label{eq:op_b5}
 {Q}^\text{Cont}_{5} &=
 [\bar{s}^a \gamma_\mu (1-\gamma_5) d^a] 
  [\bar{s}^b \gamma_\mu (1+\gamma_5) d^b],
\end{align}
where $\sigma_{\mu\nu} = [\gamma_\mu, \gamma_\nu]/2$ and $a$, $b$ are
color indices.
$Q_1$ leads to $B_K$, while $Q_{2-5}$ are the BSM operators.
The corresponding BSM B-parameters are defined as
\begin{align}
  B_i = 
  \frac{\langle \overline{K}_0 \vert Q^\text{Cont}_i \vert K_0 \rangle}
       {N_i \langle \overline{K}_0 \vert \overline{s}\gamma_5 d\vert 0 \rangle
        \langle 0 \vert \bar{s} \gamma_5 d \vert K_0 \rangle}
  \quad\quad
(N_2,\ N_3,\ N_4,\ N_5) = ( 5/3, \ 4, \ -2, \ 4/3 )\,.
\end{align}
We stress that this basis of operators differs slightly from the ``SUSY basis''
used in other lattice calculations~\cite{Boyle:2012qb,Bertone:2012cu}. 
We prefer the basis of Ref.~\cite{Buras:2000if} since we use
perturbative matching and running.

%
%

\section{Methodology and Results \label{sec:data-anal}}
%
%
\begin{table}[bp]
\caption{MILC lattices used here~\cite{Bazavov:2009bb}.
$a$ is the nominal value of the lattice spacing.
  ``ens'' and ``meas'' are the number of gauge configurations
   measurements per configuration, respectively.
  ID is an identification tag.
  \label{tab:milc-lat}}
\begin{center}
\begin{tabular}{l  l  l  c  l }
\hline
\hline
$a$ (fm) & $am_l/am_s$ & \ \ size & ens $\times$ meas  & ID \\
\hline
0.12  & 0.01/0.05    & $20^3 \times 64$  & $671 \times 9$ & C3 \\
0.09  & 0.0062/0.031 & $28^3 \times 96$  & $995 \times 9$ & F1 \\
0.06  & 0.0036/0.018 & $48^3 \times 144$ & $749 \times 9$ & S1 \\
0.045 & 0.0028/0.014 & $64^3 \times 192$ & $747 \times 1$ & U1 \\
\hline
\hline
\end{tabular}
\end{center}
\end{table}

We use the MILC lattices
listed in Table~\ref{tab:milc-lat}, setting
%
the scale using $r_1 =0.3117(6)({}^{+12}_{-31}) \textrm{ fm}$
~\cite{Bailey:2012rr}.
For the valence quarks, we use HYP-smeared 
staggered quarks~\cite{Hasenfratz:2001hp},
with parameters chosen to remove
$\mathcal{O}(a^2)$ taste-symmetry breaking at tree level.
%
Our valence $d$ and $s$ quarks have masses denoted
$m_x$ and $m_y$, respectively, 
for which we use 10 different values,
\begin{align}
\label{eq:val_q_masses}
  am_{x,y} = am_s \times {n}/{10} 
  \qquad \text{ with } n=1,2,3,\cdots,10
\,,
\end{align}
with $m_s$ is the nominal sea strange quark mass given in 
Table~\ref{tab:milc-lat}.
%
%

\begin{figure}[btp]
\begin{center}
\includegraphics[width=20pc]{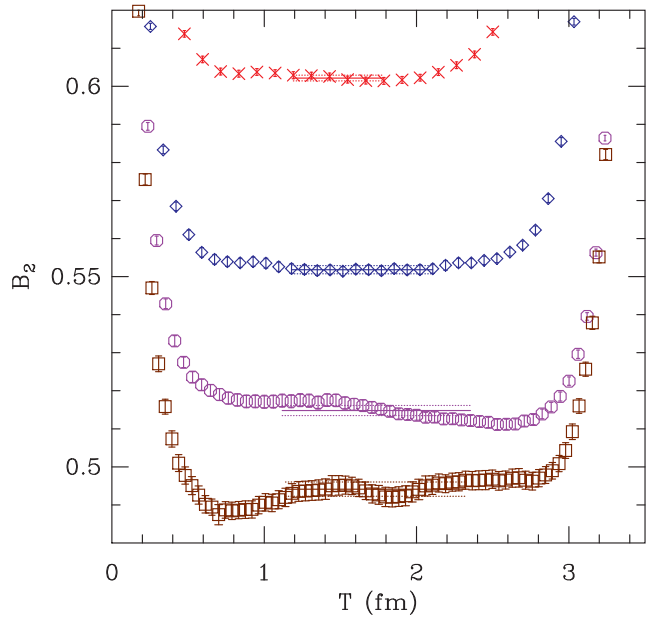}
\end{center}
\caption{$B_2(\mu=1/a)$ as a function of $T = t-t_1$.
(Red) crosses are from the coarse ensemble C3, with
$(am_x, am_y) = (0.005, 0.05)$;
(blue) diamonds are from the fine ensemble F1, with
$(am_x, am_y) = (0.003, 0.03)$;
(purple) octagons are from the superfine ensemble S1, with
$(am_x, am_y) = (0.0018, 0.018)$; and
(brown) squares are from the ultrafine ensemble U1, with
$(am_x, am_y) = (0.0014, 0.014)$.
\label{fig:b_2}}
\end{figure}

The methodology of the calculation for the BSM B-parameters is very
similar to that used for $B_K$~\cite{Bae:2010ki}. 
%
%
Many details of the lattice operators and the perturbative matching are given
in Refs.~\cite{Bailey:2012wb}, 
although some additional subtleties related to the use of the new operator
basis have led to small changes~\cite{Bailey:2012dz}. 
These, together with the renormalization
group running, will be explained in Ref.~\cite{Bailey:2013}.
The kaon and anti-kaon are produced using U(1)-noise wall-sources 
placed at timeslices $t_1$ and $t_2>t_1$, while the
four-quark operators (and bilinears needed for the B-parameters)
are placed at an intermediate time $t$.
The resulting B-parameters should be independent of $t$ when
$t$ is far enough from the wall-sources, so that
contamination from excited states is small.
Hence we fit the data to a constant in the plateau region.
The fitting range is determined using the two-point correlator from the
wall-sources to the taste-$\xi_5$ axial current.
In Fig.~\ref{fig:b_2}, we show results for $B_2$ as a function of $T =
t-t_1$ with our most physical kaon.
When fitting, we ignore the correlations between timeslices
(diagonal approximation for the covariance matrix) to avoid an
instability of the fit due to small eigenvalues of the covariance
matrix.
The fitting errors are estimated using the jackknife method.
To increase statistics, we perform multiple measurements with randomly
chosen $t_1$ on each gauge configuration (see Table~\ref{tab:milc-lat}).
We find considerable autocorrelation for the BSM $B$-parameters
on the fine, superfine, and ultrafine ensembles.
Hence we bin the data on these ensembles, using a bin size of 5.
%

%
After calculating the BSM B-parameters for 55 valence quark mass 
combinations, we perform the chiral extrapolation to the physical down 
and strange quark masses.
We first extrapolate $m_x$ to 
$m_d^\text{phys}$ at fixed $m_y$ (``X-fit''), 
and then linearly extrapolate $m_y$ to $m_s^\text{phys}$ (``Y-fit'').
In the X-fit, we fit to the form from
SU(2) staggered chiral perturbation theory (SChPT), 
which requires $m_x \ll m_y$.
Hence we take lightest four quark masses for $m_x$ 
(e.g. $m_x = \{0.005,~0.01,~0.015,~0.02\}$ on the coarse ensemble) 
and the heaviest three quark masses for the $m_y$ 
(e.g. $m_y = \{0.04,~0.045,~0.05\}$ on the coarse ensemble). 
%

For the X-fit we use the next-to-leading order (NLO)
SChPT result for the $B_j$ from
Ref.~\cite{Bailey:2012wb}, extended to higher order:
\begin{align}
\label{eq:X-fit-NNNLO}
 B_j(\text{X-fit, NNNLO}) 
  &= c_1 F_0(j) + c_2 X + c_3 X^2 + c_4 X^2 \big(\ln(X)\big)^2
  +  c_5 X^2 \ln(X) + c_6 X^3 \,.
\end{align}
Here $X = {X_P}/{\Lambda_\chi^2}$, with $X_P$
the squared mass (in physical units)
of the taste-$\xi_5$ pion composed of two light
quarks, $X_P=M^2_{xx:\text{P}}$.
For the chiral renormalization scale we take $\Lambda_\chi = 1\GeV$.
$F_0(j)$ contains the leading order and NLO chiral logarithms,
and is completely known in terms of $f_\pi$ and measured lattice pion 
masses~\cite{Bailey:2012wb}.
The $c_2$ term is the NLO analytic term.
We include three ``generic'' NNLO terms: the $c_3$ term
whis is representative of NNLO analytic terms,
and the $c_4$ and $c_5$ terms, which are representative
of NNLO chiral logarithms in continuum ChPT.
We also include one NNNLO term, with coefficient $c_6$.

We fit using the Bayesian method~\cite{Lepage:2001ym} with
parameters $c_{4-6}$ constrained to be of order unity, which is
the expectation from chiral power-counting. Specifically,
we constrain them to be $c_{4-6} = 0\pm 1$.
%
%
The full correlation matrix is included in the X-fit.

\begin{figure}[tbp]
\centering
\subfigure[X-fit]{
\label{subfig:X-fit}
\includegraphics[width=17pc]{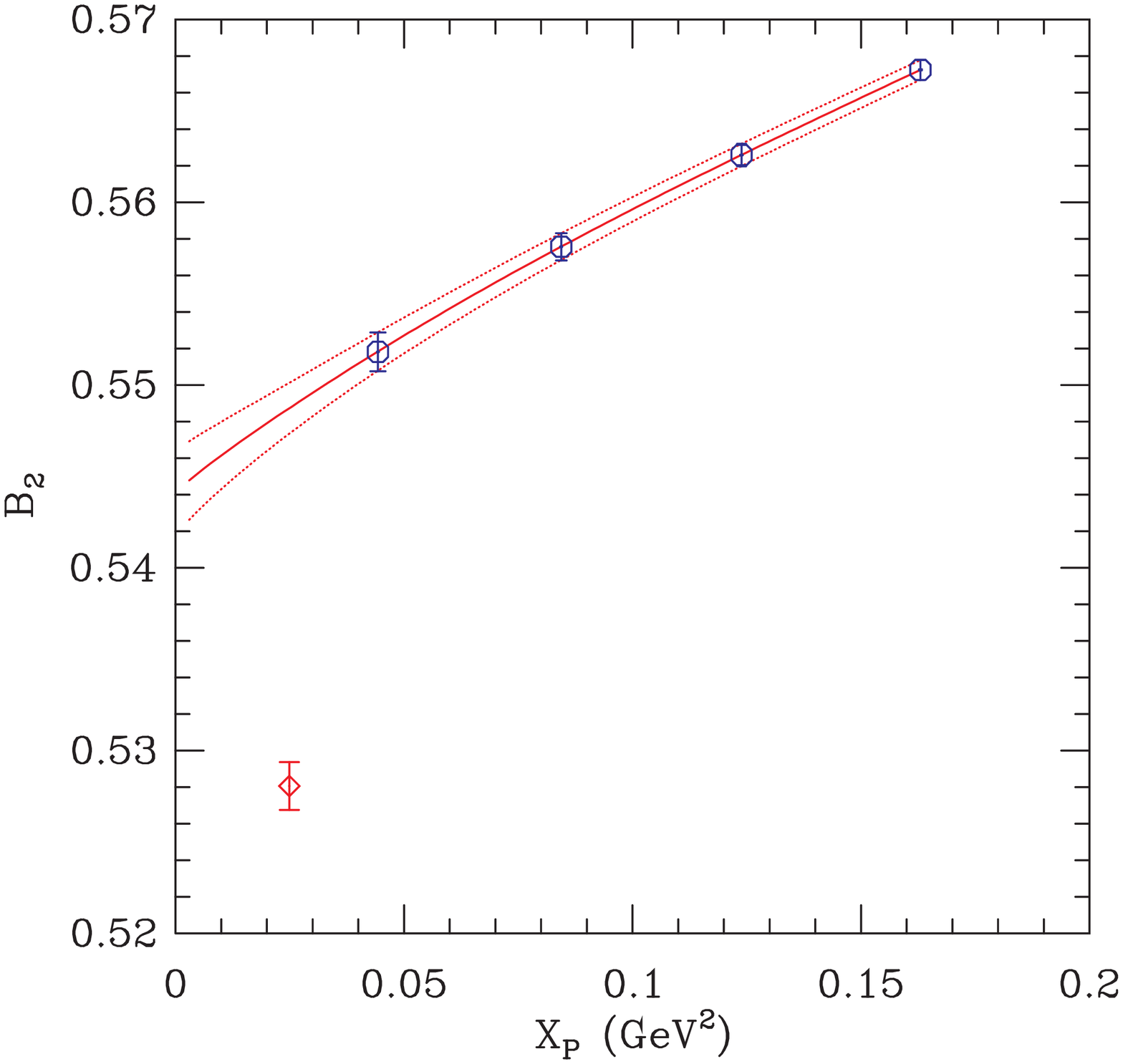}
}
\subfigure[Y-fit]{
\label{subfig:Y-fit}
\includegraphics[width=17pc]{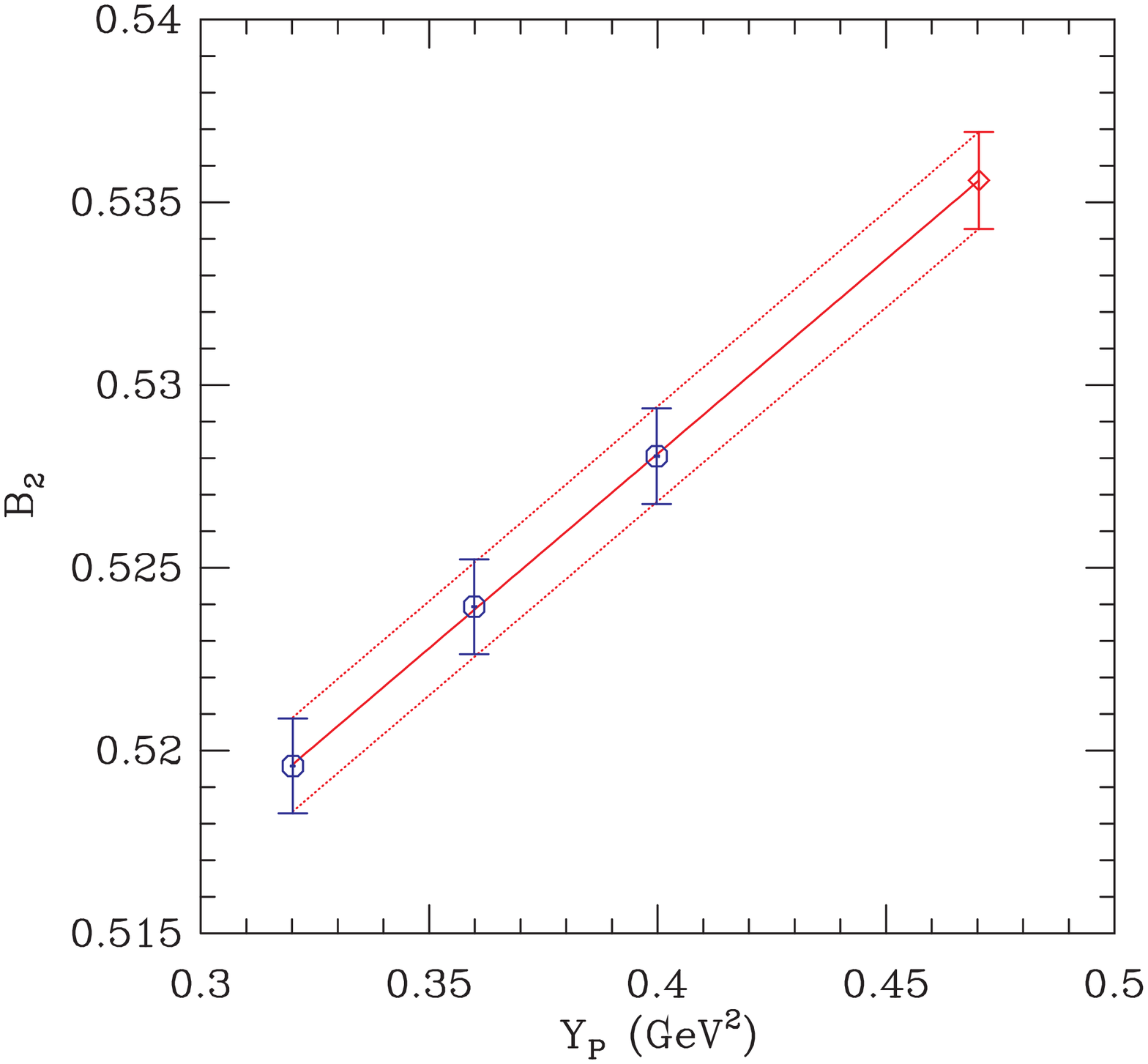}
}
\caption{(a) $B_2(\mu=1/a)$ from the NNNLO Bayesian X-fit
  vs. $X_P$, on F1, for $am_y = 0.03$.
  The red diamond represents the physical point.
  (b) $B_2(\mu=1/a)$ from the Y-fit vs. $Y_P$, on F1. 
  The red diamond corresponds to the physical strange quark mass.
  \label{fig:chiral-fit}}
\end{figure}

Having determined the parameters $c_{1-6}$, we can simultaneously
extrapolate the results to the physical
point $m_x = m_d^\text{phys}$ and remove lattice artifacts due
to taste-breaking 
in pion masses in the chiral logarithms $F_0(j)$,
as explained in Ref.~\cite{Bae:2010ki}. 
We also set $m_\ell\ne m_\ell^{\rm phys}$ in the logarithms.
In Fig.~\ref{subfig:X-fit}, we show the X-fit for $B_2$ on ensemble F1.

After the X-fit, we perform the Y-fit, in which
we extrapolate $m_y$ to the physical
strange quark mass $m_s^{\text{phys}}$.
We expect that the $B_j$ are smooth, analytic functions of
$Y_P$. It turns out that a linear form describes
the data well:
\begin{align}
  B_j(\text{Y-fit})= b_1 + b_2 Y_P
\,,
\end{align}
where $Y_P=M_{yy,P}^2$ is the squared mass of the valence pion
with composition $y\bar y$ and taste $\xi_5$.
Fig.~\ref{subfig:Y-fit} shows results of the Y-fit for $B_2$ on F1.
At this stage, we use uncorrelated fitting for the Y-fit.
%

%
After the chiral extrapolations, we know the BSM 
B-parameters evaluated at a fixed lattice spacing and matched to
the $\overline{\rm MS}$ scheme at a scale $\mu=1/a$.
In order to extrapolate to the continuum limit ($a=0$), we need to 
first run the results to a common scale $\mu$.
In the RG running, operator mixing arises in pairs:
$(Q_2^{\text{Cont}}, Q_3^{\text{Cont}})$ and $(Q_4^{\text{Cont}},
Q_5^{\text{Cont}})$.
%
%

The anomalous dimension matrix for the BSM $\Delta S=2$ operators in the
basis of Eqs.~\eqref{eq:op_b2}-\eqref{eq:op_b5} is calculated up to 
two-loop order in Ref.~\cite{Buras:2000if}. This is in the
$\overline{\rm MS}$ scheme with naive-dimensional regularization of
$\gamma_5$ and with the choice of evanescent operators made by
Ref.~\cite{Buras:2000if}.\footnote{%
A different choice of evanescent operators was made in the
one-loop matching calculation of Ref.~\cite{Kim:2011pz}. 
We have now extended
this calculation to the scheme of Ref.~\cite{Buras:2000if}.
Results will be reported in Ref.~\cite{Bailey:2013}.}
Hence we calculate the RG evolution matrix for the BSM B-parameters at
that order.
%
In the case of RG running for $B_{4,5}$, there is a removable
singularity in the standard two-loop approximate analytic solution.
To resolve this, we use the analytic
continuation method introduced in Ref.~\cite{Adams:2007tk}.
We have checked the results by numerical evolution of the RG equations.
The resulting BSM B-parameters evaluated at $\mu=2\GeV$ and $3\GeV$ 
are given in the Tables~\ref{tab:res-2gev} and \ref{tab:res-3gev}.
We note that statistical errors in the BSM B-parameters are smaller
than those in $B_K$.

\begin{table}[bp]
\caption{
  Preliminary results for BSM B-parameters and $B_K$ at $\mu=2\GeV$.
  Continuum values are obtained using linear extrapolation.
  Only statistical errors are shown.
  \label{tab:res-2gev}}
\begin{center}
\begin{tabular}{l|cccc|c}
\hline \hline
$B_j\diagdown$Lat& C3 & F1 & S1  & U1   & Continuum\\
\hline 
$B_K$         & 0.5672(52) & 0.5295(43) & 0.5362(38) & 0.5318(70) & 0.5383(66) \\ 
\hline
$B_2$         & 0.5404(09) & 0.5646(14) & 0.5967(19) & 0.6058(31) & 0.6245(30) \\
$B_3$         & 0.3689(06) & 0.4148(10) & 0.4594(14) & 0.4805(24) & 0.5032(22) \\
$B_4$         & 1.0965(23) & 1.1260(28) & 1.0911(37) & 1.0942(57) & 1.0698(56) \\
$B_5$         & 0.9278(20) & 0.9381(25) & 0.8875(31) & 0.8720(49) & 0.8432(48) \\
\hline \hline
\end{tabular}
\end{center}
\end{table}
\begin{table}[htbp]
\caption{
  Preliminary results for BSM B-parameters and $B_K$ at $\mu=3\GeV$.
  Notation as in Table~\protect\ref{tab:res-2gev}.
  \label{tab:res-3gev}}
\begin{center}
\begin{tabular}{l|cccc|c}
\hline \hline
$B_j\diagdown$Lat& C3 & F1 & S1 & U1   & Continuum\\
\hline 
$B_K$         & 0.5478(50) & 0.5114(42) & 0.5179(37) & 0.5137(67) & 0.5199(64) \\ 
\hline
$B_2$         & 0.4779(08) & 0.4993(12) & 0.5277(17) & 0.5358(28) & 0.5524(26) \\
$B_3$         & 0.3152(05) & 0.3496(08) & 0.3840(12) & 0.3997(20) & 0.4174(19) \\
$B_4$         & 1.0462(22) & 1.0750(26) & 1.0421(36) & 1.0452(55) & 1.0222(54) \\
$B_5$         & 0.9132(19) & 0.9272(24) & 0.8824(31) & 0.8714(48) & 0.8450(47) \\
\hline \hline
\end{tabular}
\end{center}
\end{table}
\begin{figure}[tbp]
\centering
\includegraphics[width=25pc]{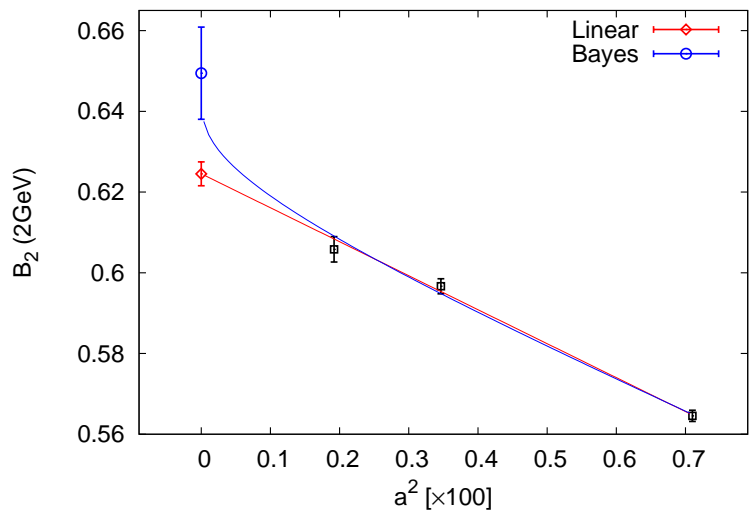}
\caption{Continuum extrapolation of $B_2$ at 2GeV.
  (Red) diamond is the result from the linear fitting function; and
  (blue) circle is the result from the Bayesian constrained fitting
  with the fitting function given in Eq.~\protect\eqref{eq:b_scaling}.
  \label{fig:conti-extrap}}
\end{figure}

The final step is the continuum extrapolation of the results.
We know that the leading
$a$ and $\alpha_s$ dependence to be~\cite{Bailey:2012dz}
\begin{align}
\label{eq:b_scaling}
 B_j = d_1 + d_2(a \Lambda)^2 + d_3 (a \Lambda)^2 \alpha_s 
    + d_4 \alpha_s^2 + d_5 (a \Lambda)^4 + \cdots
\,,
\end{align}
where $\alpha_s = \alpha_s^{\overline{\text{MS}}}(1/a)$.
We do a Bayesian fit to this form, taking the QCD scale determining 
the magnitude of discretization errors to be $\Lambda=300\text{MeV}$, 
and constraining $d_{2-5}=0 \pm 2$.
As for $B_K$, we find that fits to all four lattice spacings
are very poor, with 
$\chi^2_\text{aug}/\text{dof} = 6.6 \sim 30$ for $B_{2-5}$.
Thus we drop the results from the coarse lattice and fit to the
finest three spacings (F1, S1 and U1).
In this case, both the linear fitting (keeping only $d_1$ and $d_2$) 
and the constrained  fitting (with $d_{1-5}$) work well.
In Fig.~\ref{fig:conti-extrap}, we show an example of the continuum
extrapolation for $B_2$.
In Tables~\ref{tab:res-2gev} and \ref{tab:res-3gev}. 
we quote the results from the linear extrapolation.
Clearly the systematic errors associated with the choice of
continuum extrapolation are significantly larger than the
statistical errors.

\section{Outlook}
The next stage in our calculation is to quantify all sources of
systematic error and so draw up a complete error budget.
This requires results at other values of the light sea-quark masses
to estimate residual $m_\ell$ dependence, and at other volumes
to estimate finite volume effects. The latter can also be estimated
using SChPT, and are expected to be small. 
We also plan to investigate whether the use of ratios which cancel
chiral logarithms reduces errors in the analysis, and
to compare the results to those from an analysis using SU(3) SChPT.
We expect that, as for $B_K$, our dominant errors will  come from
to the use of one-loop matching and the continuum extrapolation.

Although our results are preliminary, it is interesting to compare
them to those found in Refs.~\cite{Boyle:2012qb,Bertone:2012cu}. 
Changing to the SUSY basis, $B_2$, $B_4$ and $B_5$ are unchanged,
while $B_3^{\rm SUSY}= (5 B_2-3 B_3)/2$. Thus our preliminary
results in the tables translate into $B_3^{\rm SUSY}=0.81$ and $0.75$
at $\mu=2$ and 3 GeV, respectively.
There are some disagreements between our results and those
of Refs.~\cite{Boyle:2012qb,Bertone:2012cu} at the 25\% level.
Determining whether these are significant will require our full error
budget.

\section*{Acknowledgments}
We thank Claude Bernard for private communications regarding the
parameters of the MILC ensembles.
C.~Jung is supported by the US DOE under contract DE-AC02-98CH10886.
 W.~Lee is supported by the Creative Research
Initiatives Program (2012-0000241) of the NRF grant funded by the
Korean government (MEST), and acknowledges
support from KISTI supercomputing
center through the strategic support program for the supercomputing
application research [No. KSC-2011-G2-06].
S.~Sharpe is supported in part by the US DOE grant
no.~DE-FG02-96ER40956.
Computations were carried out in part on QCDOC computing facilities of
the USQCD Collaboration at Brookhaven National Lab, on GPU computing
facilities at Jefferson Lab, on the DAVID GPU clusters at Seoul
National University, and on the KISTI supercomputers. The USQCD
Collaboration are funded by the Office of Science of the
U.S. Department of Energy.


\begin{thebibliography}{99}
%

\bibitem{Aubin:2009jh} C.~Aubin, J.~Laiho and R.~Van de Water,
  Phys.~Rev.~D\textbf{81}, (2010), 014507 [\texttt{arXiv:0905.3947}].

\bibitem{Aoki:2010pe} Y.~Aoki {\em et al.},
  Phys.~Rev.~D\textbf{84}, (2011), 014503 [\texttt{arXiv:1012.4178}].

\bibitem{Durr:2011ap} S.~Durr {\em et al.},
  Phys.\ Lett.\ B {\bf 705}, 477 (2011)
  [\texttt{arXiv:1106.3230 [hep-lat]}].

\bibitem{Bae:2011ff} T.~Bae {\em et al.}, SWME Collaboration,
  Phys.~Rev.~Lett.~\textbf{109}, (2012), 041601 , [\texttt{arXiv:1111.5698}].

\bibitem{Buras:2000if} A.~Buras, M.~Misiak and J.~Urban,
  Nucl.~Phys.~B\textbf{586}, (2000), 397-426 [\texttt{hep-ph/0005183}].

 \bibitem{Boyle:2012qb} 
  P.~A.~Boyle {\em et al.}, 
  Phys.\ Rev.\ D {\bf 86}, 054028 (2012)
  [{\tt arXiv:1206.5737 [hep-lat]}].

\bibitem{Bertone:2012cu} 
  V.~Bertone {\em et al.}, 
  {\tt arXiv:1207.1287 [hep-lat].}

\bibitem{Bazavov:2009bb} A.~Bazavov {\em et al.}, 
  Rev.~Mod.~Phys.~\textbf{82}, (2010), 1349 [\texttt{arXiv:0903.3598}].

\bibitem{Bailey:2012rr} J.A.~Bailey {\em et al.},
  Phys.~Phys.~D\textbf{85}, (2012), 114502 [\texttt{arXiv:1202.6346}].

\bibitem{Hasenfratz:2001hp} A.~Hasenfratz and F.~Knechtli,
  Phys.~Phys.~D\textbf{64}, (2001), 034504 [\texttt{hep-lat/0103029}].

\bibitem{Bae:2010ki} T.~Bae {\em et al.}, SWME Collaboration,
  Phys.~Phys.~D\textbf{82}, (2010), 114509 [\texttt{arXiv:1008.5179}].


\bibitem{Kim:2011pz} 
  J.~Kim {\em et al.}, 
  Phys.\ Rev.\ D {\bf 83}, 094503 (2011)
  [{\tt arXiv:1102.1774 [hep-lat]}].

\bibitem{Bailey:2012wb} J.A.~Bailey {\em et al.},
  Phys.~Phys.~D\textbf{85}, (2012), 074507 [\texttt{arXiv:1202.1570}].

\bibitem{Bailey:2012dz} 
  J.~A.~Bailey, H.~-J.~Kim, W.~Lee and S.~R.~Sharpe,
  {\tt arXiv:1210.7754 [hep-lat]}.

\bibitem{Bailey:2013} J.A.~Bailey {\em et al.},  SWME Collaboration,
  in preparation.

\bibitem{Lepage:2001ym} P.~Lepage {\em et al.},
  Phys.~Phys.~D\textbf{106}, (2002), 12-20 [\texttt{hep-lat/0110175}].


\bibitem{Adams:2007tk} D.~Adams and W.~Lee, 
  Phys.~Phys.~D\textbf{75}, (2007), 074502 [\texttt{hep-lat/0701014}].

\end{thebibliography}
\end{document}